\documentclass[prb,twocolumn,showpacs,amsmath,amssymb,superscriptaddress]{revtex4}
\usepackage{graphicx}
\usepackage{graphics}
\usepackage{dcolumn}
\usepackage{bm}
\begin{document}
\title{
Spin depolarization in the transport of holes across GaMnAs/GaAlAs/p-GaAs }
\author{L.Brey}
\affiliation{Instituto de Ciencia de Materiales de Madrid (CSIC),
Cantoblanco, 28049, Madrid, Spain.}
\author{J.Fern\'andez-Rossier}
\affiliation{Departamento de F\'{\i}sica Aplicada, Universidad de
Alicante, 03690 Alicante, Spain.}
\author{C.Tejedor}
\affiliation{Departmento de F\'{\i}sica Te\'orica de la Materia
Condensada, Universidad Aut\'onoma de Madrid, 28049 Madrid,
Spain.} \

\begin{abstract}

We study the spin polarization of tunneling holes injected from 
ferromagnetic GaMnAs into a p-doped semiconductor through a tunneling barrier. 
We find that spin-orbit interaction interaction in the barrier and in the 
drain limits severely spin injection. Spin depolarization is stronger when the 
magnetization is parallel to the current than when is perpendicular to it.

\end{abstract}

\maketitle

Achieving the injection of  spin polarized current from a ferromagnetic
material  into a semiconductor   is one of the challenges in
spintronics\cite{Ohno_book,Datta90}. However, the conductivity mismatch
between ferromagnetic metals and semiconductors prevents simple strategies of 
spin injection  in the diffusive regime \cite{Schmidt_2000,Hu}. At least two
kind of proposals have been suggested in order to circumvent this obstacle. 
First, the use of a tunnel barrier between the ferromagnetic source and the
semiconductor \cite{Rashba:2000}. Second the use of diluted magnetic
semiconductors (DMS) as a source \cite{Fiederling,Ohno:1999_b}.

Ferromagnetic diluted magnetic semiconductors materials like GaMnAs  have
raised enormous interest both because of their fundamental interest and their
potential in spintronics proposals.   One of the appealing features of  of
GaMnAs and other DMS is that they  can be integrated easily with other III-V
based heterostructures combining the magnetic and electronic functionalities.
In this direction heterostructures based in GaMnAs have been grown that
feature  strong tunneling magneto resistance effects\cite{Tanaka_Higo,Mattana}.
On the other side, the Curie Temperature is still below room temperature
although improvement in post growth annealing
techniques\cite{Potashnik:2001_,Edmonds:2002_b} in GaMnAs DMS shows the 
ability to obtain critical temperatures larger than 150K.

In GaMnAs,  Mn  act as an acceptor that supplies holes 
responsible for the long range ferromagnetic interaction between
the Mn spins\cite{Dietl:2000_,Abolfath:2001_b,Brey:2003_}. Crucial in the
understanding of the ferromagnetic phase of the material is the fact that
the spin-orbit interaction for the valence band holes 
is very strong ($\Delta \sim$ 340 meV). This large spin-orbit coupling
has several effects on the properties of  magnetic GaMnAs:
{\em i)} There is a large correlation between $T_c$ and strength of the 
spin-orbit interaction \cite{Dietl:2000_} {\em ii)} 
Spin-orbit, combined with strain
effects due to the substrate-DMS lattice mismatch, determines
the easy-axis for the magnetization\cite{Ohno:1998_a}.
{\em iii)} Spin-orbit  is also responsible for the anisotropic
magneto resistance in bulk GaMnAs\cite{Jungwirth:2002_c,Tang03}.

In this  work we address   the effect of  spin-orbit coupling on 
the injection of a spin polarized hole current from a DMS into a p-doped
paramagnetic semiconductor, via an epitaxially grown tunnel junction, {\it
i.e.}, in the coherent regime. In
particular we want to analyze how the spin polarization is
degraded, and how the spin current polarization depends on the
angle formed by the electrical current and the magnetization.
These two questions are relevant for the possible use of GaMnAs as
a source of spin polarized current. 
The system of interest consists of a ferromagnetic semiconductor
and a non magnetic semiconductor separated by a tunnel barrier. 
In particular, the left electrode is 
GaMnAs, the right electrode is p-doped GaAs and the barrier is GaAlAs. 
We consider that transport takes place by tunneling through a GaAlAs
barrier of width $d$. In this configuration spin-orbit coupling is
the same along the whole heterostructure.
We also analyze the effect produced by the quenching of the spin-orbit 
coupling only at the drain or both at the drain and the barrier.
We anticipate the main conclusions of this work:

1) Spin-orbit coupling, both at the drain and at the barrier,
significantly reduces the spin polarization of carriers injected into the non
magnetic electrode.

2) Spin injection depends significantly on the angle between the current flow
and the magnetization of the source electrode. 
When the magnetization at the source is parallel to the electrical
current, the depolarization effect is stronger than for the case
of source magnetization perpendicular to the current.

{\em Theoretical approach:} The system considered is formed by three well
defined regions along the growth direction ($z$). The \emph{left region} (L)
is the source for the spin polarized current and is formed by GaMnAs.
The \emph{barrier region} (B) is formed by GaAlAs while the
\emph{right region} (R) is a paramagnetic p-doped semiconductor,
for example Be-doped GaAs. The valence bands of this system is
described in a ${\bf k} \cdot {\bf p}$ framework by means of a
Hamiltonian having three parts:
\begin{eqnarray}
& H^L = & H_{{\bf k} \cdot {\bf p}} ^L +J_{pd} N_{Mn} S m \,
\overrightarrow{\Omega} \cdot \overrightarrow{s} \nonumber \\
& H^B = & H_{{\bf k} \cdot {\bf p}} ^B + \Delta V ^{L-B} \nonumber \\
& H^R = & H_{{\bf k} \cdot {\bf p}} ^R + \Delta V ^{L-R} .
\label{hamiltonian}
\end{eqnarray}
$H_{{\bf k} \cdot {\bf p}} ^L$, $H_{{\bf k} \cdot {\bf p}} ^B$ and
$H_{{\bf k} \cdot {\bf p}} ^R$ are six band Kohn-Luttinger
Hamiltonians for L, B and R, respectively\cite{Dietl:2000_,Abolfath:2001_b}.
Ternary compounds GaMnAs and GaAlAs are described a virtual crystal
approximation (VCA). 
We use the same Kohn-Luttinger parameters to  describe the electronic
properties of GaAs, GaMnAs and GaAlAs, i.e. $H_{{\bf k}\cdot {\bf p}} ^R=
H_{{\bf k} \cdot {\bf p}} ^L =H_{{\bf k}\cdot {\bf p}} ^B $.

In GaMnAs exchange interaction couples the spin of valence band
holes with  the spin of the Mn ions, which are randomly located in the cation
sublattice. In the mean field  and VCA \cite{Dietl:2000_}, the disordered  
exchange interaction is replaced by a homogeneous effective Zeeman field.
This approach accounts for a number of experimental observations.  The second
term of $H^L$ describes the coupling of the holes to the effective field.
There, $J_{pd}$ is the exchange coupling, $N_{Mn}$ the Mn ion density, $S$ the
spin of a Mn ion, $m$ the average polarization of the Mn spins,
$\overrightarrow{\Omega}$ the orientation of the magnetization and
$\overrightarrow{s}$ the spin of the holes. In this theoretical framework,
the ferromagnetic electrode  is characterized by the density of Mn and 
the density of holes. For a given  set of parameters in the  model
 we obtain the spin polarization of
both  Mn and holes \cite{Brey:2003_}.  

The GaAlAs barrier and p-doped GaAs drain are described by means of $k \cdot p$ 
Hamiltonians with shifts $\Delta V^{L-B}$ and $\Delta V ^ {L-R}$ with respect 
to the top of the  valence band of the ferromagnetic semiconductor.  
The precise value of the barrier height $\Delta V^{L-B}$ depends on the Al 
content in the barrier which is typically in the range between $20 \%$ and 
$40 \%$. The conduction band offset between GaAs and AlAs 
is, at the $\Gamma $ point, close to 1eV. Therefore we report 
results for an intermediate value ($30 \%$) of $\Delta V^{L-B}=300$ meV
and we have checked that results do not change qualitatively for barriers 
in the mentioned range. The shift 
$\Delta V^{L-R}$ permits to have a different carrier density in the p-doped 
region with a common Fermi energy across the heterostructure.  
Our rigid-band model neglects band-bending effects across the 
interfaces\cite{Fernandez-Rossier:2001_}.

Charge and spin transport are studied in the scattering formalism
\cite{Wessel:1989,Liu:1996,Petukhov:2002}. The quantum states of the electrodes
are described by a band index $n$ and a wave
vector ${\bf k}$,  in the framework of the 
six band ${\bf k} \cdot {\bf p}$  approximation.
These states are linear combination of $p$-like orbitals
with total angular momenta $J$=3/2 and $J$=1/2. In the presence of
spin-orbit coupling, the spin is not a good quantum number so that
the quantities conserved in the tunneling process are the energy,
$E$, and the parallel component of the wave vector, $k
_{\parallel}$ \cite{Wessel:1989,Liu:1996}. An incoming plane-wave
state from L, $| n,E,k_{\parallel};L \rangle
$, is transmitted to a plane wave $|n',E,k_{\parallel};R \rangle $
at R with a transmission amplitude $t_{n,n'}^{k_{\parallel}} (E)$.
As the group velocity in the left and right regions are in general
different, the transmission probability from a state
$|n,E,k_{\parallel};L \rangle $ to a state $|n',E,k_{\parallel};R
\rangle $ reads\cite{Wessel:1989}
\begin{equation}
T_{n,n'}^{k_{\parallel}} (E) = |t_{n,n'}^{k_{\parallel}} (E)| ^2 \,
\frac {v_{n'} ( E , k_{\parallel};R)} {v_{n} ( E ,
k_{\parallel};L)}, \label{transmission}
\end{equation}
where $v_n ( E, k _{\parallel}; L/R))$ is the group velocity,
along the $z$-direction perpendicular to the interfaces, of the
state $|n,E,k_{\parallel};L/R) \rangle $.
In our calculation,  only  incoming and transmitted states with positive group
velocity are considered. In this approach, the linear conductance of the
heterostructure can be obtained as a sum  over all transmission channels,
$G = (e^2/h) \sum _{n,n',k_{\parallel}} T_{n,n'}^{k_{\parallel}}(E_F)$.

In the following we study the degradation of the spin polarization of
carriers passing from the source (GaMnAs) to a paramagnetic drain.
We define the spin polarization of the transmitted current,
\begin{eqnarray}
\eta_{tr} \! = \! 2 \frac {\sum _{n,n',k_{\parallel}} 
 \! T_{n,n'}^{k_{\parallel}}
(E_F)  \langle n',k_{\parallel},E_F;R
\mid \! s \! \mid n',k_{\parallel},E_F;R \rangle }
{\sum _{n,n',k_{\parallel}}  \, T_{n,n'}^{k_{\parallel}} (E_F) \, }
\label{polarization}
\end{eqnarray}
where $s$ is the component of the hole spin along $\vec{\Omega}$.
We have verified that $\eta_{tr}$ along other  directions vanishes.
This quantity describes
the spin polarization of the coherently transmitted holes.
Inelastic events which can result in further spin relaxation after tunneling
are not included in our approach. 

\begin{figure}
\includegraphics[clip,width=8cm]{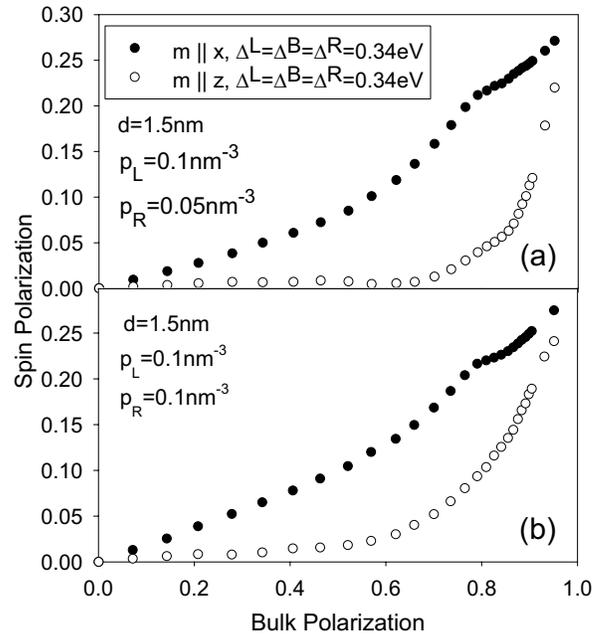}
\caption{Spin polarization $\eta _{tr}$ as a function of the bulk
polarization of GaMnAs along two different directions.} \label{fig1}
\end{figure}

{\em Results:} In Fig. \ref{fig1}, we show the spin polarization of the
transmitted current, $\eta _{tr}$, as a function of the bulk
polarization of GaMnAs, $\eta _0$.
At this point it is convenient to distinguish between the bulk 
polarization of GaMnAs, $\eta _0 $, and the polarization of the holes in the
Fermi surface of this material, $\eta _F$ \cite{Mazin99}.
The carrier density at the ferromagnetic source is fixed,
$p_L=0.1 nm^{-3}$, while two different values of $p_R$ are considered.
Two different magnetization orientation of the ferromagnetic electrode
are studied, either parallel or
perpendicular to the current flow, chosen along $z$. Results in Fig. 1,
are obtained with the same  spin-orbit coupling constant,   
$\Delta = 0.34 eV$, in all the three regions. It is notorious that $\eta_{tr}$
is significantly smaller than the bulk polarization of the
injector. The depolarization is stronger when the carriers are
polarized parallel to the current ( $z$) (open circles) 
than when they are polarized along $x$, i.e. perpendicular to the current 
(black circles). This effect is larger in the case with lower density $p_R$ 
of carriers in the p-GaAs. The feature appearing in $\eta_{tr}$ 
for $\eta _0 \simeq 0.8$, for current perpendicular to the
magnetization, coincides with the complete depopulation of 
a band of minority-spin carriers in GaMnAs. 

\begin{figure}
\includegraphics[clip,width=8cm]{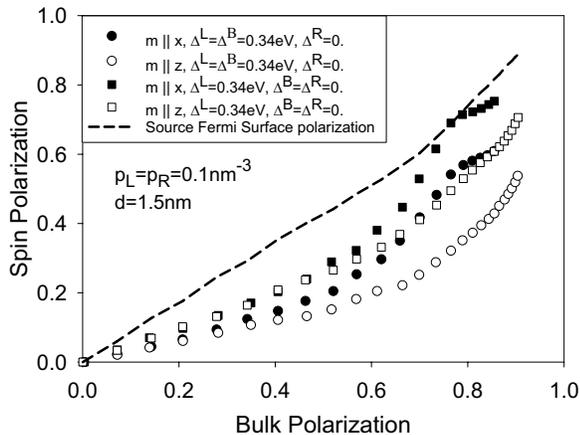}
\caption{Spin polarization $\eta _{tr}$ as a function of the bulk
polarization of GaMnAs along two different directions. Results are
for zero spin-orbit coupling only at R (circles) or at R and B (squares).
Dashed line shows the spin polarization at the Fermi level of GaMnAs.}
\label{fig2}
\end{figure}

The strong depolarization of coherently injected 
spins is produced by three mechanisms:

{\em First:} Reduction of the spin polarization at the Fermi surface.
At small bias, only electrons at the Fermi level are injected.
As it happens, the hole spin polarization at the Fermi energy $\eta _F$ is 
smaller than the bulk spin polarization $\eta _0$ (dashed line of
Fig. \ref{fig2} ). The ratio $\eta _F/\eta _0$ is roughly $0.9$ for 
$\eta _0 <0.65$ and even
larger as $\eta _0$ approaches to 1. Therefore, this effect is small in general. 

{\em Second:} Spin-orbit coupling at the barrier and the drain. 
Fig. 2 shows $\eta _{tr}$ when  spin-orbit coupling is removed either in 
the p-doped GaAs region or both in the barrier and p-doped
region.  The spin injection rate is significantly higher than the case of Fig.
1, showing that spin-orbit interaction is detrimental for successful spin
injection. This is corroborated by the fact that polarizations are larger
(lower depolarizations) when spin-orbit coupling is removed both at the barrier
and the p-doped semiconductor. As in the case of Fig. 1, depolarization is
stronger when carriers are polarized along the current direction.  The 
directional dependence is also weaker indicating that in the case of Fig. 1 it
comes from the  the spin-orbit interaction of both electrodes and barrier. 

\begin{figure}
\includegraphics[clip,width=8cm]{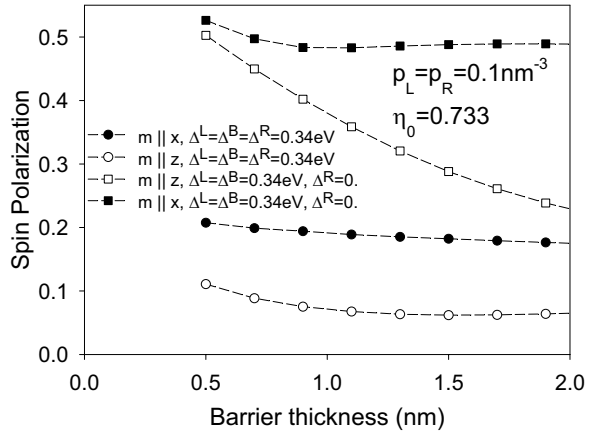}
\caption{Spin polarization $\eta _{tr}$ as a function of the
barrier width $d$ for magnetization along two different directions.}
\label{fig3}
\end{figure}

{\em Third:} Spin mixing and spin filtering in the barrier. 
Even in the absence of spin-orbit
interaction in the barrier, tunnel probability can be spin dependent.
This is known as spin filtering and accounts for the
difference between squares (spin-orbit only in the source)
and dashed line in Fig. 2. In order to clarify the effect of the barrier in the
depolarization, Fig. \ref{fig3} shows $\eta _{tr}$ as a function of the 
barrier width. The set of parameters is: $0.733$
for the polarization of GaMnAs (slightly below the kink in Fig.
2), $p_L =p_R = 0.1 nm^{-3}$. We give results for the two orientations
of the polarization as in Figs. 1 and 2, and both with and without
spin-orbit coupling at R. All the curves have the same qualitative behaviour: 
for $d=0$ the results for different orientations of $m$ must coincide. 
For increasing, but still small values of $d$, band mixing effects 
become important and the curves for $m \parallel x$ and $m \parallel z$ 
separate from each other. For further increase of $d$, these two 
curves saturate becoming flat with $d$. 

\begin{figure}
\includegraphics[clip,width=8cm]{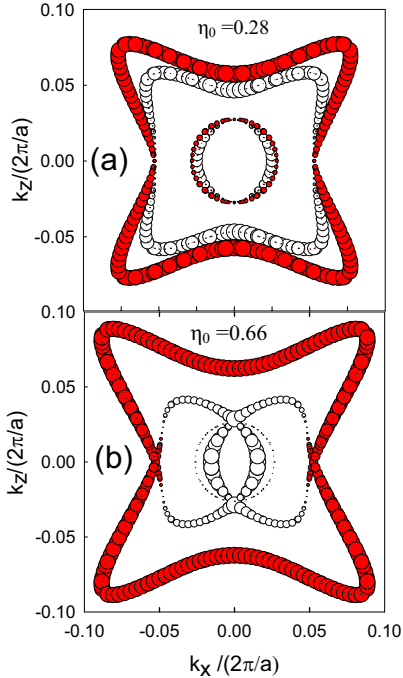}
\caption{(Color online) Expectation value of the spin component
along the direction of the polarization ($z$-direction)  in a
contour at the Fermi level of GaMnAs, for two different values of
$\eta_0$. Filled/white circles depict positive/negative direction.
The magnitude of spin is proportional to the circle
size.}\label{fig4}
\end{figure}

Let us discuss what is the physical origin of the large difference 
observed for depolarizations along the two orientations $x$ and $z$. 
In the top of the GaAs valence band, the spin-orbit coupling
creates a momentum dependent effective Zeeman field that cause the
hole angular momentum to align parallel or antiparallel to the
wavevector\cite{Yu:2001_}. This is evident in the spherical
approximation to the Luttinger Hamiltonian, where the spin-orbit
coupling is proportional to $-({\bf k} \cdot  {\bf J} ) ^2 $,
$J_i$ being the matrices for the angular momentum $3/2$. For a given ${\bf
k}$ the eigenvalues are the heavy and light bands, both with ${\bf
J}$ parallel or antiparallel to ${\bf k}$\cite{Baldereschi73}
Because of the spin-orbit coupling, the Zeeman splitting is larger
for states with ${\bf k}$ parallel to the magnetization than for states 
with ${\bf k}$ perpendicular to it, Fig. \ref{fig4}(a). In particular, for
${\bf k}$ parallel to the magnetization the heavy holes have spin
$\pm 1/2$ and an energy splitting  $J_{pd} N_{Mn} S m $, whereas
states with ${\bf k}$ perpendicular to the magnetization are
practically degenerated.
This asymmetry reflects in the tunnelling transport. For finite
spin-orbit coupling, in a tunnelling process only the energy and
the component of the wavevector perpendicular to the current is
conserved  and  states with different parallel components of the
wavevector, and different spin polarization can be mixed. This
mixing results in a lost of spin polarization in the tunnelling process.
In Fig. \ref{fig4} we see that the region of the Fermi surface, where states
with different polarization can be mixed is larger when tunnelling
current  and magnetization are parallel ($k_x$ constant) than when
they are perpendicular ($k_z$ constant).  Therefore the
degradation of the spin current is bigger in the parallel  case as
shown in all the results in Figs. 1-3, being the perpendicular
configuration the optimal for injecting spin.

In summary, spin-orbit coupling has a strong influence on the spin injection
of holes from ferromagnetic GaMnAs into p-doped GaAs via a tunneling
barrier. First of all, spin-orbit interaction reduces severely the 
efficiency of spin injection. Therefore, prospects of hole spin injections
seem better for materials with small spin-orbit like Si or GaN. Secondly, 
the spin injection rate depends on the angle between current flow and
magnetization. In particular, spin injection is significantly 
larger for samples magnetized parallel to the interfaces of the
heterostructure.  

We are indebted to G. Platero for the critical reading of the manuscript.
Work supported in part by MCYT of Spain under contract Numbers
MAT2002-04429-C03-01, MAT2002-00139, MAT2003-08109-C02-01,
Fundaci\'on Ram\'on Areces, Ramon y Cajal program and UE within
the Research Training Network COLLECT.

\bibliography{semicspin}

\end{document}